# ArXiv and the REF open access policy

Katie Shamash, Open Access Data Analyst, Jisc (Katie.Shamash@jisc.ac.uk) ORCID: 0000-0002-6827-1283

Dr Danny Kingsley, Deputy Director, Scholarly Communication, Cambridge University Library (dak45@cam.ac.uk) ORCID: 0000-0002-3636-5939

## Abstract

HEFCE's Policy for open access in the post-2014 Research Excellence Framework states "authors' outputs must have been deposited in an institutional or subject repository". There is no definition of a subject repository in the policy: however, there is a footnote stating: "Individuals depositing their outputs in a subject repository are advised to ensure that their chosen repository meets the requirements set out in this policy." The longest standing subject repository (or repository of any kind) is arXiv.org, established in 1991. arXiv is an open access repository of scientific research available to authors and researchers worldwide and acts as a scholarly communications forum informed and guided by scientists. Content held on arXiv is free to the end user and researchers can deposit their content freely. As of April 2018, arXiv held over 1,377,000 eprints.

In some disciplines arXiv is considered essential to the sharing and publication of research. The HEFCE requirements on repositories are defined in the Information and Audit Requirements which lists the "Accepted date", the "Version of deposited file" and "available open access immediately after the publisher embargo" are expected as part of the REF submission.

However, while many records in arXiv have multiple versions of work, the Author's Accepted Manuscript is not identified and there is no field to record the acceptance date of the work. Because arXiv does not capture these two specific information points it does not meet the technical requirements to be a compliant subject repository for the purposes of REF. This paper is presenting the case that articles deposited to arXiv are, in general, compliant with the requirements of the HEFCE policy. The paper summarises some work undertaken by Jisc to establish if there are other factors that can indicate the likelihood of formal compliance to the HEFCE policy.

# Contents

# Introduction

The Higher Education Funding Council for England (HEFCE)'s Policy for open access in the post-2014 Research Excellence Framework states "authors' outputs must have been deposited in an institutional or subject repository". (HEFCE, 2015) There is no definition of a subject repository in the policy: however, there is a footnote stating: "Individuals depositing their outputs in a subject repository are advised to ensure that their chosen repository meets the requirements set out in this policy."

The longest standing subject repository (or repository of any kind) is arXiv.org, established in 1991 by Paul Ginsparg at the Los Alamos Laboratory. It is currently hosted at Cornell University. arXiv is an open access repository of scientific research available to authors and researchers worldwide and acts as a scholarly communications forum informed and guided by scientists. Content held on arXiv is free to the end user and researchers can deposit their content freely.

As of April 2018, arXiv held over 1,377,000 e-prints in Physics, Mathematics, Computer Science, Quantitative Biology, Quantitative Finance, Statistics, Electrical Engineering and Systems Science, and Economics. In some disciplines arXiv is considered essential to the sharing and publication of research.

The HEFCE requirements on repositories are defined in the Information and Audit Requirements (HEFCE, 2014) which lists the "Accepted date", the "Version of deposited file" and "available open access immediately after the publisher embargo" are expected as part of the REF submission.

This means there are two barriers to using the papers deposited in arXiv for compliance with REF's open access policy:

- Many records in arXiv have multiple versions of work attached, however these are classified as Version 1, Version 2 rather than using the NISO terminology widely accepted in research, of Submitted Version, Author's Accepted Manuscript and Version of Record (NISO, 2008).
- There is no field to record the acceptance date of the work. The HEFCE policy requires this information because compliance is tied to a deposit within a three month period from acceptance.

Because arXiv does not capture these two specific information points it does not meet the technical requirements to be a compliant subject repository for the purposes of REF.

During 2015-2016 a significant effort was made between the UK Higher Education sector and arXiv to negotiate undertaking the necessary development work to add these data fields to arXiv records. Technical specifications were agreed and in principle agreement for funding of the work was secured. However personnel challenges resulted in an inability for this project to meet the HEFCE policy start date deadline of 1 April 2016.

This paper is presenting the case ***that articles deposited to arXiv are, in general, compliant with the requirements of the HEFCE policy***. The paper summarises some work undertaken by Jisc to establish if there are other factors that can indicate the likelihood of formal compliance to the HEFCE policy.

# 2. Study 1 - Estimating the formal compliance of articles on arXiv with REF policy

## 2.1 Study background

For a work to be formally compliant with the HEFCE policy, it must:

1. Have metadata deposited in repository within three months of acceptance
2. Have author accepted manuscript or published version deposited within three months of acceptance
3. Be made open access immediately after the publisher embargo

It is not possible to assess what percentage of works on arXiv formally comply with REF due to the following missing information:

1. Acceptance date
2. Information about version available on arXiv

Instead, three factors were studied that could indicate the likelihood of formal compliance: presence of a DOI, date of metadata deposit, and date of last update in arXiv.

## 2.2 Study Method

A study was undertaken by Jisc in 2016 considering a sample of articles available in arXiv taken from 2011-2015 to establish whether it is possible to determine which version of the work is uploaded and when. The arXiv API was used to download metadata in XML format for a sample of 1200 articles submitted to arXiv between 2011 and 2016. The XML was parsed and, where a DOI was present, the Crossref API was called on the DOI and the metadata was joined with arXiv. Basic analysis was then performed using Python scripts and Excel.

## 2.3 Presence of a DOI

The presence of a DOI is the most basic element of compliance, since without it, there is no sure way of identifying the published article corresponding to the pre-print. In addition, CrossRef can be used to supplement the metadata about the article for a more complete bibliographic record.

Digital Object Identifiers (DOI) are allocated to articles by publishers. Records within arXiv that contain a DOI indicate the work has been published and the author has updated the record to include the DOI. Roughly half of the sample of articles (53%) listed a DOI. The presence of a DOI is the most basic element of compliance, since without it, there is no sure way of identifying the published article corresponding to the pre-print. In addition, CrossRef can be used to supplement the metadata about the article for a more complete bibliographic record.

As is indicated in Figure 1, DOIs are present for 58% of the sampled articles added to arXiv in 2011, and for 60% of articles added to arXiv in 2012. The percentage then begins to decline—likely because articles deposited to arXiv in more recent years have had less time to be accepted and published in a journal.

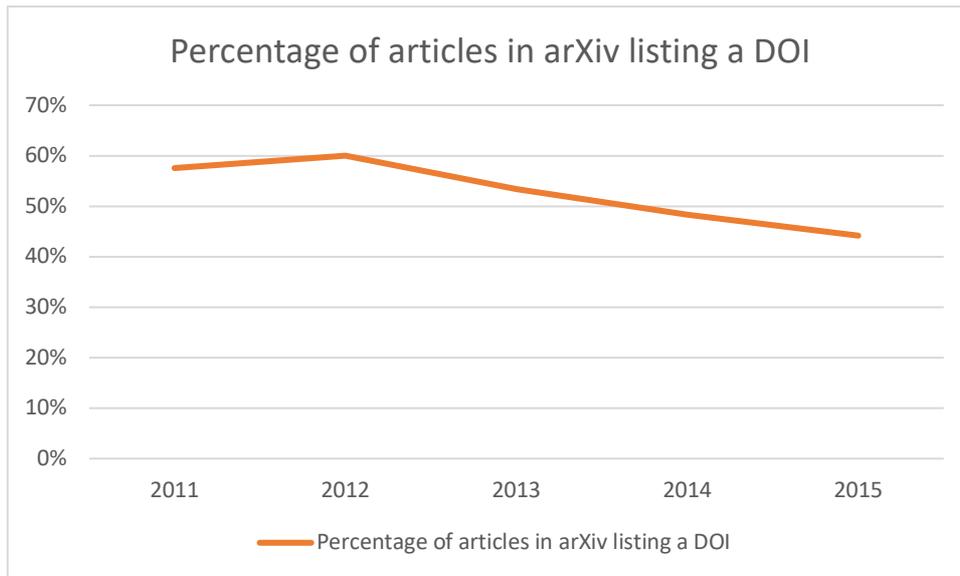

*Figure 1 Percentage of articles on arXiv listing a DOI by year. Overall, 53% of articles published 2011-2015 listed a DOI.*

The high level of DOIs indicates that authors are returning to their records to update the information after publication.

## Date of metadata deposit to arXiv

The study considered the question of the date of metadata deposit to arXiv. For those articles with a DOI listed, nearly all (89%) had been uploaded to arXiv before the date of journal publication. On average, articles were deposited in arXiv 146 days before publication in a journal. This is graphically demonstrated in Figure 2.

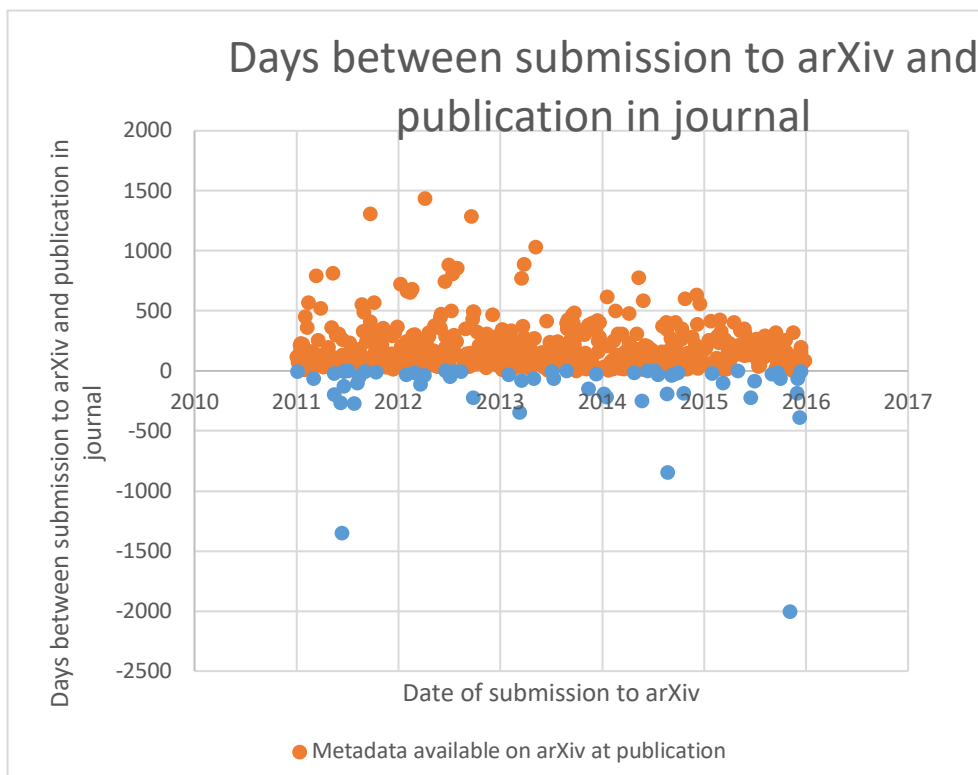

*Figure 2 Days between submission to arXiv and publication in journal as a function of time. The vast majority of published articles on arXiv are submitted well before publication in a journal.*

## 2.5 Version available on arXiv

For an article to be REF compliant, the deposited version must be the Author's Accepted Manuscript or the final Version of Record. While it is possible for a record in arXiv to have multiple versions of the work, these are identified as Version1, Version 2 and so on. There is no distinction between the Submitted Version, the Author's Accepted Manuscript or the final Version of Record. Therefore, the only way to definitively ascertain whether the version deposited to arXiv would correspond with REF policy is to do a full-text analysis, which is beyond the scope of this study.

Instead, as a proxy, the date of last update on arXiv was compared with the date of publication. If the text of the article was updated close to the date of publication, it would indicate that the author had updated the text after the manuscript had been accepted. While by no means a reliable measure, this metric provides an indication of whether authors are updating the text of their article once the manuscript is in its accepted form.

As the figure below shows, most articles on arXiv are updated 90 days prior to publication or later. On average, arXiv versions are updated two months before the date of publication.

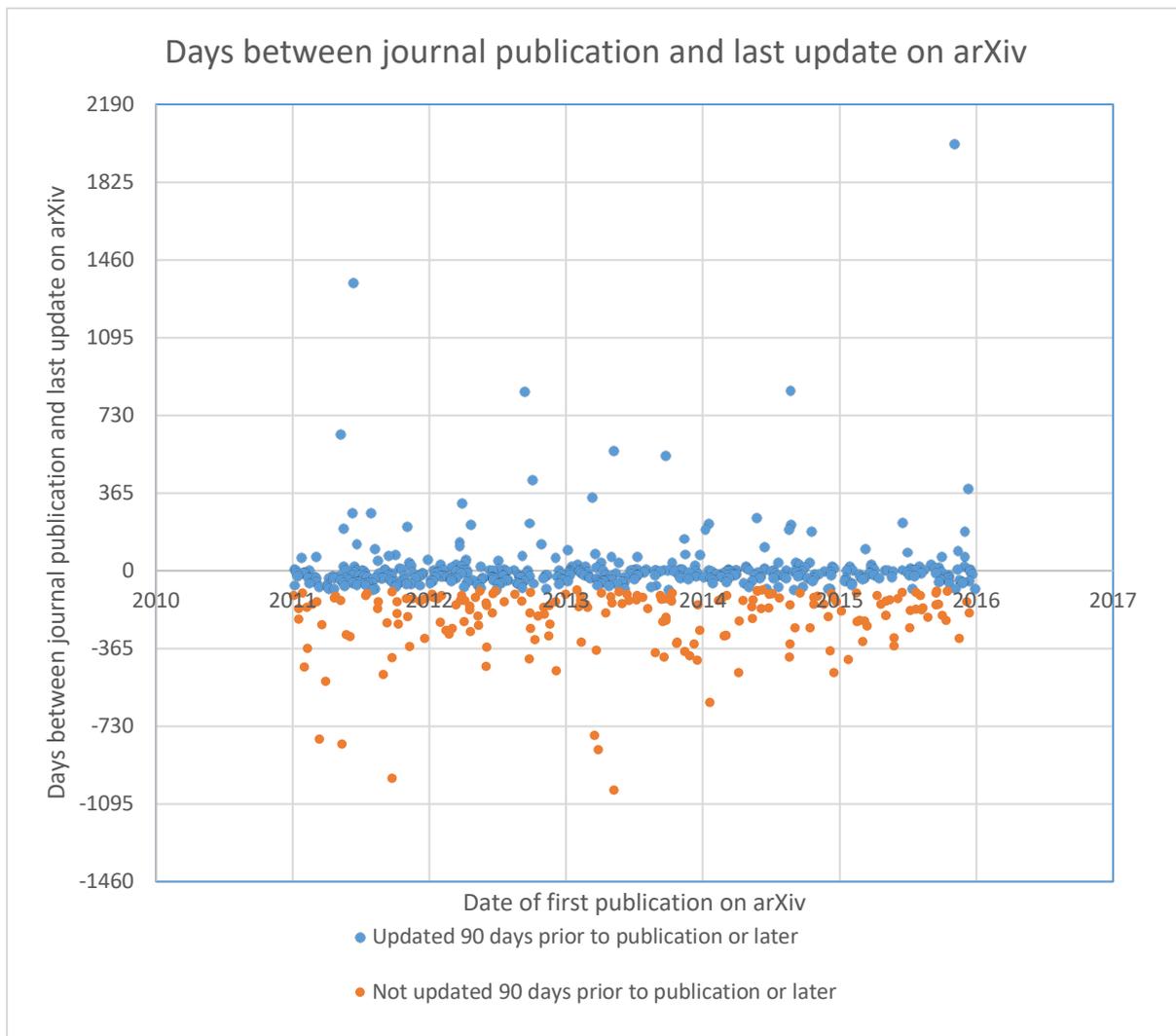

*Figure 3 Days between journal publication and last date of update on arXiv as a function of time. Most articles on arXiv are updated 90 days prior to publication or later.*

If most articles are updated three months prior to publication or later, this indicates that researchers in these disciplines are updating their work shortly after the time of acceptance.

## 2.6 Discussion

The study findings that articles are on average uploads two months before publication have been echoed in other studies. For example, the findings are consistent with Klein et. al.'s (2016) finding that 90% of published articles are available on arXiv before publication. This large percentage is not surprising given arXiv's role as a repository for preprints. In addition, the findings align with Larivière et. al.'s (2013) finding that preprints are likely to be published in a journal within a year of publication on arXiv. Two-thirds of articles were updated three months prior to publication or later. The average time between acceptance and publication is 5.72 months for physics journals and 5.11 months for mathematics journals (Björk and Solomon 2013).

This study has demonstrated that for those articles in arXiv with a DOI (and therefore, by logical extension, published) have been uploaded within two months of publication. This is a very strong indicator that the peer reviewed and corrected version of the work (the AAM) has been uploaded.

Even where the version available on arXiv does not correspond to the Author's Accepted Manuscript or final Version of Record, the differences are likely to be so small as to be insignificant. A large scale text analysis of arXiv articles found that "the vast majority of final published versions are largely indistinguishable from their pre-print versions" (Klein et. al. 2016). Even the earliest possible pre-print versions showed insignificant differences compared to their published counterparts. ArXiv articles then likely meet the spirit, if not the letter, of the REF policy in respect to version.

# 3. Study 2 - Comparison of usage of articles in arXiv with those in journals

## 3.1 Literature review on arXiv usage

### 3.1.1 Coverage

ArXiv makes an enormous amount of research available open access. An analysis of topics within arXiv by Hu et al (2015) found that arXiv's coverage of nearly all topics is growing much faster than journal coverage of those same topics. This shows arXiv's increasing popularity as a medium for disseminating research and its high potential for growth.

Figure 4 breaks down by discipline the over 100,000 submissions to arXiv per year.

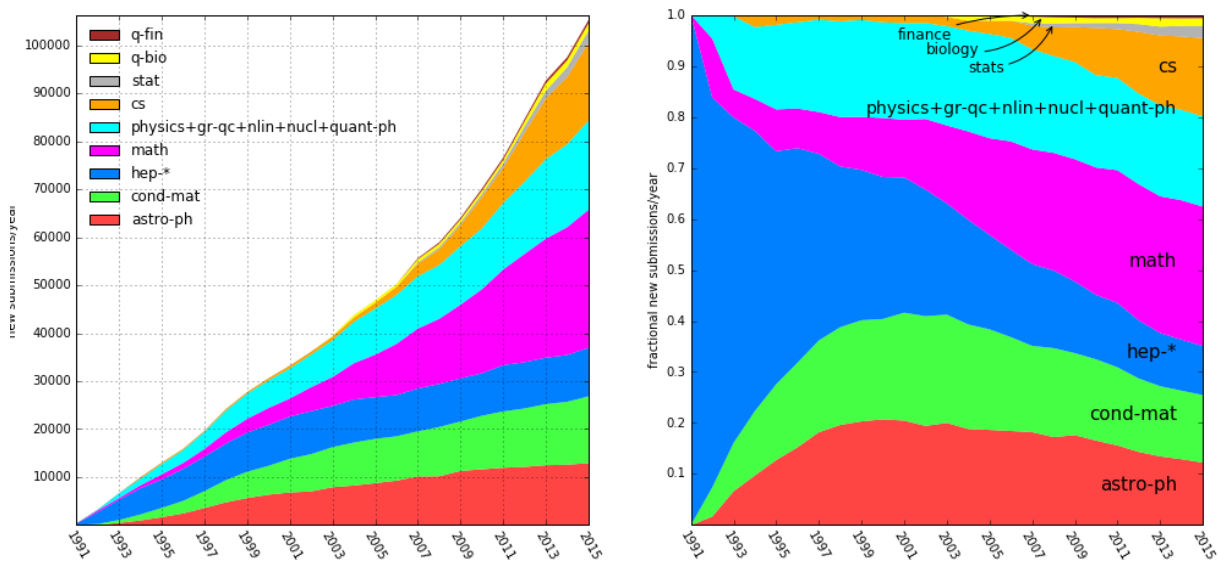

*Figure 4 Number of submissions to arXiv by discipline and percentage of submissions by discipline. Source: arxiv.org, reprinted with permission*

A total of 3.6% of Web of Science indexed papers from 2010 in all disciplines were available on arXiv in 2013 (Larivière et. al. 2013). The fields with the highest concentration of published papers available on arXiv are mathematics (21%), physics (20%), and earth and space (12%) (Larivière et. al. 2013). Within certain subfields of these disciplines, coverage is much higher. Nearly all articles published in the main High Energy Physics (HEP) journals are also available in some version on arXiv: a 2009 paper found that yearly between 1999 and 2008, more than 90% of articles published in five major HEP journals also appeared on arXiv (Gentil-Beccot, Mele and Brooks, 2009).

### 3.1.2 Citations and downloads

ArXiv is a platform for discovery and use of research. As of February 2016, arXiv had 719,637,748 total downloads. Figure 6 demonstrates the increase in the number of downloads over time. ArXiv currently receives over 10 million downloads a month (arXiv, 2016).

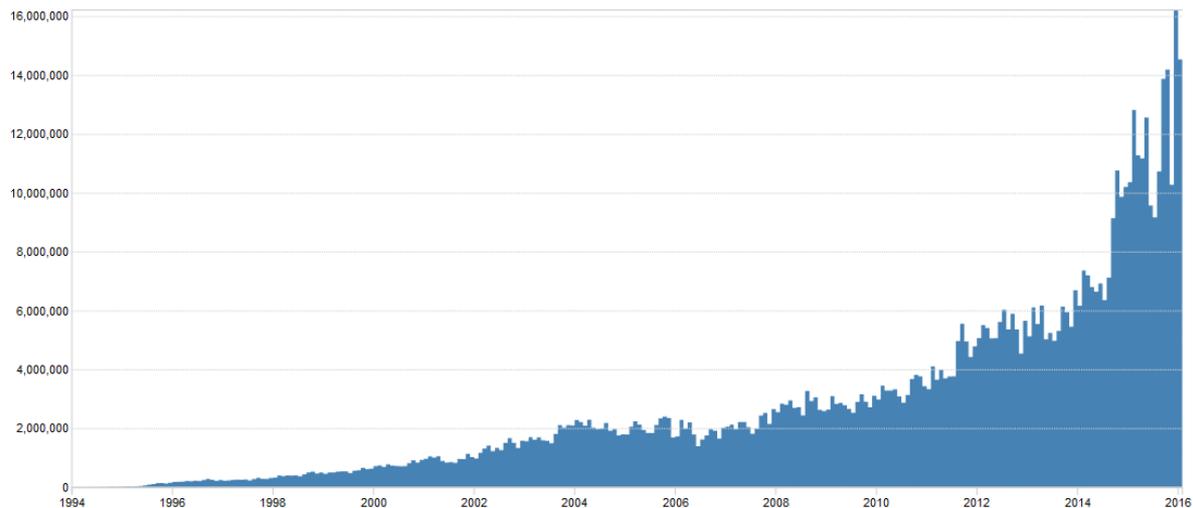

*Figure 5 - ArXiv monthly download rate. Source: arXiv.org, reprinted with permission*

### 3.1.3 Citations of papers in arXiv

Researchers are increasingly citing arXiv as a source. Work undertaken by Aman (2015) demonstrates an increasing number of published papers are citing arXiv e-prints, with over 7,000 papers in Web of Science citing an e-print in arXiv in 2013.

Within journals, an increasing proportion of published papers reference arXiv. The UK is the third most likely country to cite arXiv, with 30,776 of UK-authored papers citing arXiv, from 1991-2013, 10.6% of all cited references to arXiv came from the UK .

ArXiv is primarily cited as a source when the article has not yet been published in a journal. Citations peak in the year after publication on arXiv: half of these citations to arXiv refer to e-prints from the previous year (Aman 2015).

Citations to arXiv vary across disciplines. Nuclear and particle physics is the field that is most likely to cite arXiv: in this discipline, 7% of cited references between 2007 and 2011 referred to arXiv (Larivière et. al. 2013).

The high rate of citation to arXiv articles suggests that within those disciplines where arXiv is used, eprints are an accepted source of academic research (Aman 2015).

### 3.1.4 Citation advantage for arXiv papers

Not only are arXiv papers widely used and widely cited in and of themselves, depositing an e-print on arXiv creates a significant citation advantage for the published version of the paper. High Energy Physics (HEP) articles in arXiv accumulate citations prior to publication, and have an average impact factor nearly double that of articles not present in arXiv. The citation advantage extends beyond HEP: articles from four mathematics journals studied received 35% more citations if they were deposited to arXiv (Davis and Fromerth, 2006).

Research suggests that the citation advantage is due to earlier availability and a quality differential, rather than to open access (Moed 2006). HEP articles published in open access journals do not exhibit any citation advantage compared to articles published in closed access journals (Gentil-Beccot, Mele, and Brooks, 2009). This may indicate that the citation advantage to articles available on arXiv is not due to their being open access, but to their being available earlier and discoverable more easily. Other reasons behind the citation advantage may be that authors are more likely to deposit research on arXiv if they perceive it as high impact, or that authors who know to use arXiv publish higher impact papers. Regardless of the reason, the presence of a citation advantage indicates that arXiv is central to the way researchers in particular disciplines discover, share, and disseminate research.

A study compared the levels of citations between articles available only in arXiv, those published and not available in arXiv, and those both published and available in arXiv. The advantage for articles both published and available in arXiv is large: since 1997, this group of articles has had an 'impact factor' hovering on or above 5 (calculated as the number of citations in that year over the number of works published in the past two years). Meanwhile, articles published but not submitted to arXiv had an impact factor that steadily decreased from 1.4 to below 1 in the same period: (Gentil-Beccot, Mele, and Brooks, 2009).

ArXiv has high usage and a higher rate of citation compared to other established subject specific repositories. ArXiv documents received on average 2.5 citations per document, compared to 1.7 for Social Science Research Network (SSRN), 1.3 for Research Papers in Economics (RePEC), and 1.1 for PubMed Central (PMC) (Li, Thelwall and Kousha, 2015). A comparison can be seen in Figure 11. Disciplinary norms can explain some, but not all, of these differences; for example, arXiv receives more citations per document than PubMed Central, even though biomedical papers tend to contain more citations than physics papers (Anon. 2011).

The number of citations to arXiv is increasing at a rate comparable or greater to that of other subject specific repositories REPEC, SSRN, and PMC (Li, Thelwall, and Kousha, 2015). ArXiv is widely used across all disciplines. While the majority of citations to arXiv come from science and mathematics papers, other subjects (Arts and Humanities, Social Sciences, and Medicine), are increasingly citing cite arXiv too (Li, Thelwall, and Kousha, 2015).

Looking at citation count alone likely understates actual usage of arXiv, as researchers are encouraged to reference the published article if it is available (Larivière et. al., 2013). Download statistics suggest that readers are highly engaged with arXiv. Data from the high energy physics literature database INSPIRE (formally SPIRES) showed that if a record returned a link to a both a e-print on arXiv and a published article on a journal website, 82% of users would click to arXiv (Gentil-Beccot, Mele, and Brooks, 2009). This suggests that readers prefer to read articles on arXiv even when the published version is available. Similarly, articles available on arXiv receive 23% fewer downloads on the publisher's website two years following publication (Davis and Fromerth, 2007). This may reflect readers' preference for finding the article on arXiv. Shuai, Pepe and Bollen (2012) found that while downloads peak in the months following submission to arXiv, they remain steady in the six months following publication. [1]

## 3.2 Study Method

The analysis compared institution-level journal and arXiv usage for journals with high rates of coverage in arXiv.

This analysis drew on data from the Journal Usage Statistics Portal (JUSP) and on arXiv institution level usage data for 2015. It is difficult to compare the two data sources accurately due to limitations in the data: ArXiv usage data only gives a total number of downloads per institution per year, and JUSP data gives a total number of downloads per journal per year.

---

[1] By contrast, Henneken at. al. found that arXiv downloads of astronomy articles dropped to nearly zero following the publication of the article in a journal (2007).

Since the data is not available at an article level from either source, it is difficult to discern what proportion of the JUSP usage can be attributed to arXiv content, and conversely, what proportion of arXiv usage can be attributed to journal content. As such, any comparisons between the two are only broad, order of magnitude comparisons.

First, a sample of 1200 articles was taken from arXiv for 2010-2015. A five year period was chosen because 92% of usage is within five years after publication (Nicholas et. al. 2005). The DOIs found on arXiv were then looked up in CrossRef for journal information. With this, it was possible to determine the five most popular journals that preprints on arXiv are later published to. These journals were *Physical Review Letters*, *Monthly Notices of the Royal Astronomical Society*, *Physical Review A*, *Physical Review E*, *Physical Review B*, and *Physical Review D*. Using the Crossref API, metadata for a sample of 100 DOIs from each of these popular journals was downloaded, and these DOIs were looked up in arXiv to determine journal coverage.

2015 JUSP data was downloaded for these journals for the UK institutions with the highest usage of arXiv in 2015 according to arXiv usage statistics. The institutions are: Imperial College London, University of Cambridge, University of Edinburgh, and University of Oxford.

These datasets were joined and calculated with a rough estimate of usage as follows:

*Institution's journal usage of content available on arXiv = Institution's total usage of journal x percentage of journal content available on arXiv*

*Institution arXiv usage of content available in journal = Institution's total usage of arXiv x percentage of content on arXiv available in journal*

This method assumes that usage is evenly distributed among articles and journals, which is unlikely to be the case. Articles published in major journals may receive more usage on arXiv as they have higher visibility. Articles not available on arXiv may receive higher journal usage as this is the only way to access them; or, they may receive lower usage if they are low impact articles which have not been submitted to arXiv for that reason.

Using a DOI as a source of identification contributes to error as some works on arXiv may not list the DOI of their published version. Furthermore, arXiv's API has a known bug for DOI searches where some DOIs are not retrieved, even if they are available (arXiv API forum, 2015). Because of this bug, some journals did not return any results on the arXiv API, and so were excluded from the study. For those which were found, arXiv coverage was low compared to Gentil-Beccot, Mele and Brooks' estimate of nearly 100% coverage (2009). This suggests that arXiv usage is likely underestimated.

### 3.2 Results

These sources of error aside, estimated arXiv usage was consistently close, and in some cases higher than, estimated journal usage. Overall, arXiv usage was estimated at 39% of total usage, as demonstrated in Figure 12. The data breakdown for this Figure is included as Appendix 1.

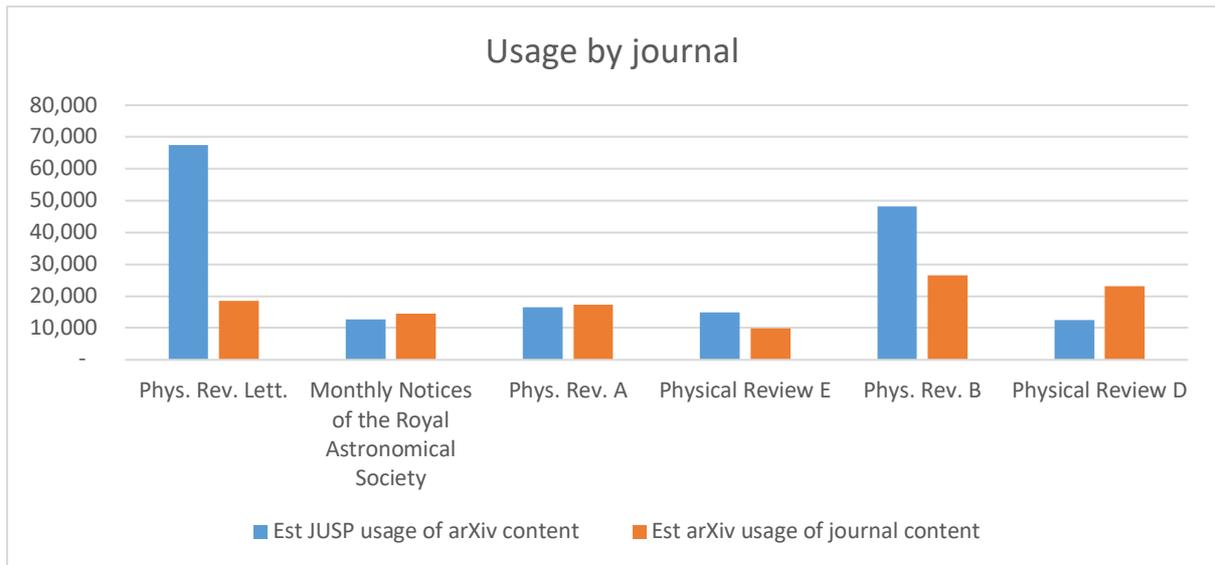

*Figure 12 Estimated JUSP and arXiv usage by journal*

At an institutional level, arXiv usage was consistently between 30-50% of estimated total usage, and is shown in Figure 13. The data breakdown for this Figure is included as Appendix 1.

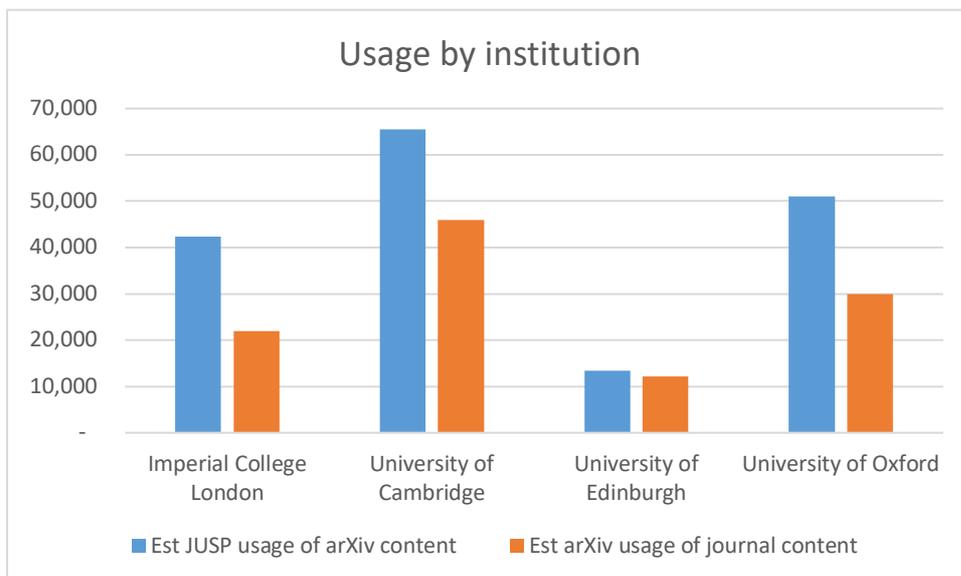

*Figure 13 Estimated JUSP and arXiv usage by institution*

## 3.3 Discussion

Previous research has shown that arXiv's coverage of nearly all topics is growing much faster than journal coverage of those same topics. The fields with the highest concentration of published papers available on arXiv are mathematics, physics and earth & space. Nearly all articles published in the main High Energy Physics (HEP) journals are also available in some version on arXiv. The arXiv repository currently receives over 10 million downloads a month.

Researchers are increasingly citing arXiv as a source, primarily when the article has not yet been published in a journal. Citations to arXiv vary across disciplines, with it being highest in the discipline of nuclear and particle physics. Indeed depositing an e-print on arXiv creates a significant citation advantage for the published version of the paper.

ArXiv has high usage and a higher rate of citation compared to other established subject specific repositories. Looking at citation count alone likely understates actual usage of arXiv, as researchers are encouraged to reference the published article if it is available. However, the presence of a citation advantage indicates that arXiv is central to the way researchers in particular disciplines discover, share, and disseminate research.

This new study has demonstrated estimated usage of arXiv by the academic community was consistently close, and in some cases higher than, estimated journal usage. This shows that academics are using arXiv as a source for research and discovery in a similar way to their use of journals. It is clear that to this community, at least, the material placed in arXiv is 'good enough' to be used as a source of information and papers in their research.

# 4. Study 3 - Responses from academics about their use of arXiv and relevance to their discipline

## 4.1 Study Method

To establish researchers' attitude towards arXiv, in October 2016 an email was sent to Cambridge University academics from relevant disciplines to ask about their usage of arXiv. The six academics who responded were a mix of early career and senior researchers and professors across the disciplines of astrophysics, pure mathematics, high energy physics, theoretical physics, and condensed matter physics.

They were asked the following questions:

1. What is arXiv to you?

2. How do you use it?

3. Do you go anywhere else to find material in your field, and if so where do you go?

## 4.2 Results

Nearly all of the respondents (n=5) reported using arXiv frequently to discover new literature. Three reported checking the website daily, while another reported checking 3 times a week. One called arXiv "the first port of call to search for literature on a given subject, and access away from institutional access to journals." Another said, "The arXiv is extremely important for me. It is the main tool that I use to obtain a broad view of current research in my general area… Often I tell my colleagues about things I have seen on the arXiv." They regarded arXiv as an important tool for discovery.

Researchers also valued arXiv for access (n=3). "The arXiv provides me with… a repository on which I can get free access to almost any paper," said an early career researcher. "I… find arXiv extremely valuable for access to paywalled papers when I am off campus," said a senior researcher. "It is much easier to simply go to arXiv and download what I want, without having to jump through multiple sites and remember login details."

All researchers reported using other methods to find articles (n=6), such as google search (n=3), subject sites (n=2), library sites (n=1), and Research Gate (n=1). However, none of the researchers mentioned institutional repositories as a substitute for arXiv. No researchers mentioned using journal sites directly to find research either.

Although we didn't ask directly, many of the researchers also mentioned using arXiv to publish their own work (n=3). "I post all my published and some unpublished work (e.g., lecture notes; review/pedagogical notes) that I consider of use to the community on the arXiv," said one researcher. "In my community... you will probably find that this attitude towards the arXiv is quite typical, if not systematic."  One reported updating preprints with the accepted version after peer review, but noted that others would only post an article following acceptance. However, another warned that papers "perhaps might not be updated that reliably, so if a paper has been out for a few years getting the final journal version is better, or sometimes the version on the author's website. "

### 4.2 Discussion

Overall, the interviews revealed an enthusiasm for arXiv and a desire for funders to recognize its importance to their field. "arXiv is great, it does everything I want, and if funders / universities accepted it as meeting open access requirements it would be perfect!" one researcher enthused.

Another said, "I only wish [arXiv's] use was made compulsory by funding agencies. If it were systematic, I would have the certainty that all published work can be searched there, with hyperlinks to published journals as appropriate." The clear sentiment was that arXiv's importance to their field should be recognized and encouraged.

Tellingly, one comment was: "I post accepted papers to institutional open access repositories as this is a requirement of funding agencies. But aside from this, I never use these sites". This comment points to the heart of the issue about excluding arXiv from being a compliant repository. Researchers who use arXiv are being forced to do double the work to maintain compliance. In many cases they are simply ignoring the policy and not uploading their work to their institutional repository.

## 5. Conclusion

This paper has made the case: *that articles deposited to arXiv are, in general, compliant with the requirements of the HEFCE policy*. It sought to answer the question through three approaches.

The first study considered alternative measures for formal compliance with REF policy and demonstrated that for those articles in arXiv with a DOI, most have been uploaded within two months of publication. This is a very strong indicator that the peer reviewed and corrected version of the work (the AAM) has been uploaded. In any case, it should be noted that the first two years of the HEFCE policy (1 April 2016 – 30 March 2018) has required deposit within three months from publication, and the recent HEFCE survey is asking the community whether this should be the ongoing policy. If this were to become the case, then this block to using arXiv.org for compliance is removed.

Even where the version available on arXiv does not correspond to the Author's Accepted Manuscript or final Version of Record, other large scale research has shown the differences are likely to be so small as to be insignificant. ArXiv articles then likely meet the spirit, if not the letter, of the REF policy in respect to version.

The second approach looked at the literature into arXiv usage and ran a comparison study of arXiv with journal usage. It is clear that arXiv has excellent coverage of some disciplines, and is very highly used, with over 10 million downloads a month. Researchers are increasingly citing arXiv as a source.

Placing an e-print on arXiv creates a significant citation advantage for the published version of the paper. ArXiv has high usage and a higher rate of citation compared to other established subject specific repositories. The new study demonstrated estimated usage of arXiv by the academic community was consistently close, and in some cases higher than, estimated journal usage. This shows that academics are using arXiv as a source for research and discovery in a similar way to their use of journals. It is clear that to this community, at least, the material placed in arXiv is 'good enough' to be used as a source of information and papers in their research.

The final approach was a qualitative one, asking the opinion of a small number of researchers who use arXiv. Overall, the interviews revealed an enthusiasm for arXiv and a desire for funders to recognize its importance to their field. One commenter noted how little they use institutional repositories. Researchers who use arXiv are being forced to do double the work to maintain compliance. In many cases they are simply ignoring the policy and not uploading their work to their institutional repository.

This paper clearly demonstrates the academic community in some disciplines continue to make the vast majority of their work openly accessible through arXiv.org and consider this to be an essential aspect of their research dissemination. There is an argument that the end goal of the HEFCE policy - providing open access to UK research outputs – is being achieved in those disciplines using arXiv.

# Appendix 1: Tabular data underpinning journal usage analysis

| Journal | Total JUSP usage | Percentage of articles in arXiv from journal | Percentage of articles in journal on arXiv | Estimated JUSP usage of arXiv content | Estimated arXiv usage of journal content | arXiv usage / total usage |
|---|---|---|---|---|---|---|
| Physical Review Letters | 103,771 | 3% | 65% | 67,451 | 18,507 | 22% |
| Monthly Notices of the Royal Astronomical Society | 25,780 | 2% | 49% | 12,632 | 14,458 | 53% |
| Physical Review A | 23,661 | 3% | 70% | 16,563 | 17,350 | 51% |
| Physical Review E | 27,745 | 1% | 54% | 14,982 | 9,832 | 40% |
| Physical Review B | 77,822 | 4% | 62% | 48,250 | 26,603 | 36% |
| Physical Review D | 13,509 | 3% | 92% | 12,428 | 23,133 | 65% |
| **Total for all eight journals** | **272,288** | **16%** | **65%** | **172,306** | **109,883** | **39%** |

*Table 1 JUSP and arXiv usage and coverage for the selected journals.*

| Institution | Total JUSP usage | Estimated JUSP usage of arXiv content | Estimated arXiv usage of journal content | arXiv usage / total usage |
|---|---|---|---|---|
| Imperial College London | 65,726 | 42,344 | 21,909 | 34% |
| University of Cambridge | 104,614 | 65,545 | 45,938 | 41% |
| University of Edinburgh | 21,085 | 13,374 | 12,142 | 48% |
| University of Oxford | 80,863 | 51,043 | 29,895 | 37% |
| **Total for all four institutions** | **272,288** | **172,306** | **109,883** | **39%** |

*Table 2 JUSP and arXiv usage for the selected institutions.*

# Appendix 2: Informal arXiv survey – questions and responses

1. What is arXiv to you?
2. How do you use it?
3. Do you go anywhere else to find material in your field, and if so where do you go?

## A2.1 Senior researcher, astrophysics:

'To answer your first question, arXiv is an incredible aid to my work, and my productivity would be significantly lower without it.

I (and my collaborators) deposit all our publications on arXiv. Personally, I tend to deposit a preprint version on submission to a journal, and then update with the accepted version after refereeing. Others will only post a paper to the arXiv after acceptance.

As posting to arXiv is ubiquitous in astronomy, it is very easy for me to keep abreast of the field but reading the "new submissions" page for astrophysics each morning.

I also find arXiv extremely valuable for access to paywalled papers when I am off campus. I know I can login to raven, or access paywalled journals though the library website. But it is much easier to simply go to arXiv and download what I want, without having to jump through multiple sites and remember login details.

I post accepted papers to institutional open access repositories as this is a requirement of funding agencies. But aside from this, I never use these sites. The same goes for journal websites - I go to these to submit my own papers, but I do not access any content through them.

I search for papers using the SAO/NASA astrophysics data system (ADS; http://adsabs.harvard.edu/abstract_service.html). ADS gives a direct link to the pdf from the journal, and also a link to the arXiv posting or each paper.

tl;dr version - arXiv is great, it does everything I want, and if funders / universities accepted it as meeting open access requirements it would be perfect!'

## A2.2 Early career researcher, pure mathematics:

### >1. What is arXiv to you?

A very important repository of work online - in my field most papers are posted there first, and very few papers written today are never posted there. They perhaps might not be updated that reliably, so if a paper has been out for a few years getting the final journal version is better, or sometimes the version on the author's website, but the arXiv is an essential place for keeping up with research in my field.

### > 2. How do you use it?

I typically check the arXiv each morning for new papers in my field. I put up my own papers there (or one of my co-authors will), and often find papers there when I'm looking for references if they have been published within the last say 10 years.

### > 3. Do you go anywhere else to find material in your field, and if so > where do you go?

Webpages of academics, and webpages of journals primarily. Occasionally books and very infrequently old print journals.'

### A2.3 Early career researcher, high energy physics:

>1. What is arXiv to you?

A data base to post and find (not necessarily peer reviewed) papers.

> 2. How do you use it?

Post papers and read papers.

> 3. Do you go anywhere else to find material in your field, and if so > where do you go?

University library sites, Inspire website, Research Gate

I would add that the arXiv's moderation has run into a considerable amount of criticism over the last year or two. (eg Nature article on the matter) Moderators have been known to overstep their censoring authority in several cases, imposing their personal opinions on what types of research are allowed to be made public.'

### A2.4 Professor, theoretical physics:

> 1. What is arXiv to you?

The arXiv is extremely important for me. It is the main tool that I use to obtain a broad view of current research in my general area.

> 2. How do you use it?

I look at new papers on the arXiv often, probably about 3 times a week. Often I tell my colleagues about things I have seen on the arXiv.

> 3. Do you go anywhere else to find material in your field, and if so where do you go?

I find more specialised information from going to meetings and from word-of-mouth. If I am looking at a topic in depth I will go back to the earlier literature. Recent papers on the arXiv may prompt me to look at a topic in more depth.

### A2.5 Senior researcher, theoretical physics:

>1. What is arXiv to you?

A repository of all work published (and some unpublished).

>2. How do you use it?

I post all my published and some unpublished work (e.g., lecture notes; review/pedagogical notes) that I consider of use to the community on the arXiv. [In my community (theoretical condensed matter physics), you will probably find that this attitude towards the arXiv is quite typical, if not systematic.]

The first port of call to search for literature on a given subject, and access away from institutional access to journals. I only wish its use was made compulsory by funding agencies. If it were systematic, I would have the certainty that all published work can be searched there, with hyperlinks to published journals as appropriate.

>3. Do you go anywhere else to find material in your field, and if so where do you go?

The arXiv has a time limit (it was only created relatively recently), and some members of the physics community at large (in particular experimentalists publishing in glossy journals with an embargo policy) do not post all their work on the arXiv. For this reason, after an initial search of the literature on the arXiv, it is sometimes necessary to look elsewhere. Most typically, a google search is sufficient to provide links to journal articles pre-dating (or sadly not posted on) the arXiv.'

### A2.6 Early career researcher, condensed matter physics:

#### >1. What is arXiv to you?
The arXiv provides me with a list of the most recent papers in my field, and a repository on which I can get free access to almost any paper

#### >2. How do you use it?
I read the daily feed of new manuscripts, and search for historical papers

#### >3. Do you go anywhere else to find material in your field, and if so where do you go?
To find papers I often use a Google search. To be exposed to recent breakthroughs I also look at other specialist news feeds.